\documentstyle[12pt]{article} 
\begin{document}
\newcommand{\beq}{\begin{equation}}
\newcommand{\eeq}{\end{equation}}
\newcommand{\beqn}{\begin{eqnarray}}
\newcommand{\eeqn}{\end{eqnarray}}
\newcommand{\slp}{\raise.15ex\hbox{$/$}\kern-.57em\hbox{$\partial
$}}
\newcommand{\slA}{\raise.15ex\hbox{$/$}\kern-.57em\hbox{$A$}}
\newcommand{\lnA}{\raise.15ex\hbox{$/$}\kern-.57em\hbox{$A_{c}^{N}$}}
\newcommand{\slB}{\raise.15ex\hbox{$/$}\kern-.57em\hbox{$B$}}
\newcommand{\bP}{\bar{\Psi}}
\newcommand{\bC}{\bar{\chi}}
 \newcommand{\hs}{\hspace*{0.6cm}}

\title{ Factored coset approach to bosonization in the context of topological
backgrounds and massive fermions}
\author{ 
M.V.Man\'{\i}as$^{a,b}$,C.M.Na\'on$^{a,b}$, and M.L.Trobo$^{a,b}$}
\date{March 1996}
\maketitle

\def\thepage{\protect\raisebox{0ex}{\ } La Plata 96-03}
\thispagestyle{headings}
\markright{\thepage}

\begin{abstract}

\hs We consider a recently proposed approach to bosonization in which the original
fermionic partition function is expressed as a product of a  $G/G$-coset model
and a bosonic piece that contains the dynamics. In particular we show how the
method works when topological backgrounds are taken into account. We also 
discuss the application of this technique to the case of massive fermions.

\end{abstract}
 
\vspace{3cm}
Pacs: \\ 
\hspace*{1,7 cm} 11.10.-z \\
\hspace*{1,7 cm} 11.15.-q

\noindent --------------------------------

\noindent $^a$ {\footnotesize Depto. de F\'\i sica.  Universidad
Nacional de La Plata.  CC 67, 1900 La Plata, Argentina.}

\noindent $^b$ {\footnotesize Consejo Nacional de Investigaciones
Cient\'\i ficas y T\'ecnicas, Argentina.}

\newpage
\pagenumbering{arabic}

\hs In a recent paper \cite{Ni}, Theron et al  presented an alternative 
approach to bosonization in two dimensions using the path integral formalism.
They obtained a complete derivation of the bosonization dictionary for both
the Abelian and non-Abelian cases. In the Abelian case they started with 
the generating functional for current-current correlation functions of 
free Dirac fermions in two dimensional Euclidean space:

\beq
Z= \int D\bP~ D\Psi DA \exp \{- \int d^2x~ [\bP i\slp\Psi - j_{\mu} A_{\mu}]\}
\label{1}
\eeq

\noindent where $j_{\mu} = \bP \gamma_{\mu} \Psi$.\\ 

\hs In a general gauge, that is without fixing the gauge, the method consists 
in making a gauge transformation in the fermionic variables which gives rise 
to a delta of conservation of the fermionic current $j_{\mu}=\bP \gamma_{\mu}
\Psi$. Appropiately representing $\delta(\partial_{\mu}j_{\mu})$ and making a 
chiral change of variables $Z$ can be factorized in 
terms of a $G/G$-coset fermionic partition function, and a bosonic part which 
contains the dynamics:

\beq
Z= \int D\bP ~ D\Psi ~ DB_{\mu} e^{- S_{cf}} \int D\phi DA_{\mu}
e^{-S_{bos}[ \phi, A_{\mu} ]} \label{2}
\eeq

\noindent where $S_{cf}$ is the action of the fermionic coset $U(1)/U(1)$
model.\\
\hs This is one of the main achievements of the new approach. Indeed, in the
standard decoupling technique of the path integral bosonization \cite{Ga},
$Z[A_{\mu}]$ is expressed as a bosonic partition function multiplied by
the vacuum to vacuum amplitude of free fermions. In the framework of 
Ref.\cite{Ni} 
the fermionic factor corresponds to constrained fermions, which are dynamically 
trivial (in the sense that both fermionic currents are set equal to zero).
Thus, the bosonizing character of the procedure becomes more apparent.\\
\hs Taking this into account it is interesting to analyze the applicability of 
this new method to the study of other physical situations in which the bosonization 
procedure is known to be more involved than in the case of free massless fermions.
In this work we focus our attention on two of such situations. Firstly we 
consider a model of fermions coupled to a vector field $A_{\mu}$, allowing 
this field to carry a non-trivial topological charge: $\oint A_{\mu} dx^{\mu}=-2\pi N$
\cite{BC}. One of the more interesting features of this model is the existence 
of the so called minimal correlation functions which, being zero for trivial 
topology, develop non-zero values when $N \neq 0$.\\ 
\hs Finally we go back to trivial
topology and briefly show how to extend the method of \cite{Ni} to the case 
in which fermions are massive. This is an important point since much of the 
pioneering work on bosonization was done in the context of massive models
\cite{Col}.\\
\hs On the other hand, our discussion concerning this matter could also be 
helpful in order to use the ideas of \cite{Ni} in the study of 2D 
statistical-mechanics models away from criticality. In this context $\bP \Psi$
is the energy density of the system and $m \propto (T-T_{c})$.\\
\hs Let us start with the generating functional introduced in Ref \cite{BC}:

\beq
Z= \sum_{N} \int D\bP ~ D\Psi DA_{\mu} \exp \{-\int d^2x~ {\bP [i\slp + \slA] 
\Psi}\}
\label{3}                                                                          
\eeq

with $A_{\mu}=A_{\mu}^{c(N)}+a_{\mu}$, where $A_{\mu}^{c(N)}$ is a fixed
(classical) configuration with topological charge $N$:

\beq
\oint A_{\mu}^{c(N)} dx^{\mu}=-2\pi N
\label{4}
\eeq

while $a_{\mu}$ stays in the topologically trivial ($N=0$) sector.\\

\hs Following the procedure developed in  \cite{Ni} to obtain a coset model
in a general gauge, we perform a gauge trasformation

\beq
\Psi \rightarrow e^{i\eta(x)} \Psi \hspace{2cm}
\bP \rightarrow \bP e^{-i\eta(x)}
\label{5}
\eeq

in the generating functional (\ref{3}). We then have

\beq
Z= \sum_{N} \int D\bP ~ D\Psi DA_{\mu} \exp \{-\int d^2x~ {[\bP i\slp \Psi + 
j_{\mu}A_{\mu} + (\partial_{\mu}j_{\mu})\eta]} \}
\label{6}                                                                          
\eeq

Due to gauge invariance, this partition function is  $\eta$-independent. Then 
we can integrate over $\eta$ and obtain

\beq
Z= \sum_{N} \int D\bP ~ D\Psi DA_{\mu}  \delta(\partial_{\mu}j_{\mu})
\exp \{-\int d^2x~ {[\bP i\slp \Psi + 
j_{\mu}A_{\mu}]} \}
\label{7}                                                                          
\eeq

Now we represent $\delta(\partial_{\mu}j_{\mu})$ in the form

\beq
\delta(\partial_{\mu}j_{\mu})= \int DB_{\mu} D\theta \exp \{-\int d^2x~ 
{[B_{\mu}j_{\mu} + \frac{1}{\pi} B_{\mu} \epsilon_{\mu\nu} \partial_{\nu}
\theta]} \}
\label{8}
\eeq

Note that there is no factor of $i$ in the exponent of the above expression
just because we are representing the delta functional in Euclidean space.\\
Introducing this expression in the generating functional and making the 
shift $B'_{\mu}=B_{\mu}+a_{\mu}$, which Jacobian is equal to one, we obtain

\beqn
Z & = & \sum_{N} \int D\bP ~ D\Psi DA_{\mu} DB'_{\mu} D\theta
\exp \{-\int d^2x~ [\bP i\slp \Psi   
 +  j_{\mu} ( A^{c(N)}_{\mu} + B'_{\mu}) + \nonumber\\ 
& + & \frac{1}{\pi} B'_{\mu} \epsilon_{\mu\nu} \partial_{\nu} \theta\} 
- \frac{1}{\pi} a_{\mu} \epsilon_{\mu\nu} \partial_{\nu} \theta] \} 
\label{9}                                                                          
\eeqn

In order to factorize out a constrained fermionic model we have to eliminate
the linear term in $B'_{\mu}$. To this end we make a chiral transformation
with parameter $\sigma$ in the fermionic variables. This change yields a
Fujikawa Jacobian given by \cite{Fu}: 

\beq
J_{F}= \exp \{ \frac{1}{\pi} \epsilon_{\mu\nu} \partial_{\nu} \sigma (B'_{\mu}
+A^{c(N)}_{\mu}) - \frac{1}{2\pi}\sigma \Box \sigma \} 
\label{10}                                                                          
\eeq

The generating functional becomes:

\beq
Z= \sum_{N}\int D\bP ~ D\Psi DA_{\mu} DB'_{\mu} D\theta e^{-S} 
\label{11}                                                                          
\eeq

where

\beqn
S & = & \int d^2x \{ \bP ~i\slp \Psi + (B'_{\mu}+A^{c(N)}_{\mu}
+ \epsilon_{\mu\nu}\partial_{\nu}\sigma)j_{\mu}+  
\frac{1}{\pi} \epsilon_{\mu\nu} \partial_{\nu} \theta B'_{\mu}
- \frac{1}{\pi} \epsilon_{\mu\nu} \partial_{\nu} \theta a_{\mu} \nonumber\\ 
&-& \frac{1}{\pi} \epsilon_{\mu\nu} \partial_{\nu} \sigma  A^{c(N)}_{\mu}  
-\frac{1}{\pi} \epsilon_{\mu\nu} \partial_{\nu} \sigma B'_{\mu} +
\frac{1}{2\pi}\sigma \Box \sigma \} 
\label{12}
\eeqn

\hs Now choosing $\sigma=\theta$, we cancel the linear term in $B'_{\mu}$
and obtain

\beq
Z= \sum_{N}\int D\bP ~ D\Psi DA_{\mu} DB'_{\mu} D\sigma e^{-S'} 
\label{13}                                                                          
\eeq

with

\beq
S  =  \int d^2x \{ \bP ~i\slp \Psi + (B'_{\mu}+A^{c(N)}_{\mu}
+ \epsilon_{\mu\nu}\partial_{\nu}\sigma)j_{\mu}  
- \frac{1}{\pi} \epsilon_{\mu\nu} \partial_{\nu} \sigma A_{\mu} 
+\frac{1}{2\pi}\sigma \Box \sigma \} 
\label{14}
\eeq

In order to write this generating functional as the product of a 
$\frac{U(1)}{U(1)}$ fermionic coset model and a bosonic part, like in
Ref.\cite{Ni}, we make the shift $B'_{\mu} \rightarrow B'_{\mu}-
\epsilon_{\mu\nu}\partial_{\nu}\sigma$. Thus we have

\beq
Z= \sum_{N}\int D\bP~ D\Psi DB'_{\mu} e^{-S_{cf}^{(N)}} \int D\sigma DA_{\mu}
e^{-S_{bos}[\sigma,A_{\mu}^{(N)}]}
\label{15}                                                                          
\eeq

where

\beq
S_{cf}^{(N)}=\int d^{2}x \bP~[i\slp + \slB '+ \slA^{c(N)}] \Psi
\label{16}
\eeq

and 

\beq
S_{bos}[\sigma,A_{\mu}]=\int d^{2}x \frac{1}{\pi} [\sigma \frac{\Box}{2} 
\sigma - \epsilon_{\mu\nu} \partial_{\nu} \sigma A_{\mu}]
\label{17}
\eeq

\hs Let us stress that in our case both the $\frac{U(1)}{U(1)}$ fermionic 
coset factor and the bosonic one have a non-trivial topological structure.\\
\hs If we want to identify this result with the one obtained in Ref.\cite{BC}
we have to make a second chiral transformation in the fermionic variables, 
in order to decouple the $B_{\mu}$ field from the fermions \cite{Ga}. In 
this way we recover a factorized generating functional $ Z = Z_{Fer} \times 
Z_{Bos}$, with  $Z_{Fer}$ having the same zero modes problem which was 
studied in Ref.\cite{BC}.\\ 
\hs Calling the parameter of the transformation $\phi$, we obtain:

\beqn
Z & = & \sum_{N} \int ~ D\bC~ D\chi  \exp \{-\int d^2x ~ [\bC (i\slp + \lnA) 
\chi]\} \times  \nonumber\\
& \int & DA_{\mu} D\sigma D\phi  \exp \{\frac{1}{\pi} \int d^2x 
[\epsilon_{\mu \nu} \partial_{\nu} \sigma  A_{\mu} - \frac{1}{2} \sigma 
\Box \sigma + \epsilon_{\mu \nu} \partial_{\nu} \phi  A_{\mu}^{c} + 
\frac{1}{2} \phi \Box \phi]\} \nonumber \\ 
& = & Z_{Fer} \times Z_{Bos} \label{18} 
\eeqn

\hs From this expression, one can conclude that only the $N = 0$ sector contributes to the 
$Z$, a well established result for  massless fermions \cite{BC}.\\ 
\hs Now, in order to compute fermionic correlation functions we have to add 
source terms in the form:

\beq
Z_{\rho_{R}, \rho_{L}} =  
 \int D\bP~ D\Psi DA_{\mu} \exp \{-\int d^2x~ [\bP i\slp\Psi - j_{\mu} A_{\mu} +
 \rho_{R} \bP_{R} \Psi_{R} + \rho_{L} \bP_{L} \Psi_{L}]\}
\label{19}
\eeq

\noindent so that, differentiating $n_{R}$ times ($n_{L}$) with respect to $\rho_{R}$ 
($\rho_{L}$) we obtain the minimal correlation function in the $n_{R}$ 
($n_{L}$) sector, and making the cross derivatives $n_{R}$ times with respect 
to $\rho_{R}$ and $n_{L}$ times with respect to $\rho_{L}$ we arrive to the 
non-minimal correlator in the $\mid n_{R} - n_{L} \mid$ topological sector.\\

\hs Let us consider the minimal functions in the $n_{R}>0$ sector (equivalent
results are obtained in the $n_{L}<0$ sector) as follows:

\beq
<\prod_{i=0}^{n_{R}-1} \bP_{R} \Psi_{R} (x_{i}) > = Z^{-1}   
\frac{\delta^{n_{R}} Z}{ \delta \rho_{R} (x_{0})....\delta \rho_{R} 
(x_{n_{R} -1})} \label{20}
\eeq

\hs This expression can be factorized as the product of a fermionic mean value
and two bosonic parts in the form

\beqn
<\prod_{i=0}^{n_{R}-1} \bP_{R} \Psi_{R} (x_{i}) > & = &
<\prod_{i=0}^{n_{R}-1} \bC_{R} \chi_{R} (x_{i})>_{fer}
<e^{\sum_{i=0}^{n_{R}-1} \beta_{i} \phi (x_{i})} >_{bos}  \nonumber\\
& < & e^{\sum_{i=0}^{n_{R}-1} \beta_{i} \sigma (x_{i})} >_{bos}  
\label{21}
\eeqn

where $\beta_{i}=2$ for all values of $i$, and

\beq
L_{F}= \bP~[i\slp + \slA^{c(N)}] \Psi
\label{22}
\eeq

\beq
L_{\sigma}= \frac{1}{\pi} [\sigma \frac{\Box}{2} \sigma-
\epsilon_{\mu\nu} \partial_{\nu} \sigma A_{\mu}]
\label{23}
\eeq
and

\beq
L_{\phi}= -\frac{1}{\pi} [\phi \frac{\Box}{2} \phi +
\epsilon_{\mu\nu} \partial_{\nu} \phi A_{\mu}]
\label{24}
\eeq

\hs For simplicity we shall consider a distribution of the
topological charge given by

\beq
\epsilon_{\mu \nu} \partial_{\mu} A_{\nu}^{C} = \sum_{i=0}^{n_{R}-1}
\alpha_{i}(x) \delta^{2}(x - x_{i}) \label{25}
\eeq

with
\beq
\sum_{i=0}^{n_{R}-1}\alpha_{i}(x_{i}) = - 2 \pi N \label{26}
\eeq

\hs This distribution corresponds to a Nielsen-Olesen configuration \cite{NO},
in which the vortex width tends to zero (see Ref.\cite{mnt} for details).\\
\hs For the fermionic part we obtain

\beq
F(x_{0},....,x_{n_{R}-1}) = det_{0}( i\slp + \lnA) e^{ \sum_{i=0}^{n_{R}-1}
\alpha_{i} \beta_{j} \Box^{-1}_{x_{i}, x_{j}}} \prod_{i<j} \mid x_{i} - x_{j} 
\mid^{2}
\label{27}
\eeq

while the bosonic piece (i.e. the product of the last two factors in (\ref{16}))
is given by

\beq
B(x_{0},....,x_{n_{R}-1}) = e^{- \sum_{i,j} ( \frac{\alpha_{i} \alpha_{j}}
{2 \pi} + \alpha_{i} \beta_{j} + \frac{ \pi}{2} \beta_{i} \beta_{j})
\Box^{-1}_{x_{i}, x_{j}}}   \label{28}
\eeq
\hs So, the minimal correlation function for the $n_{R} >0$ topological sector
reads:

\beq
<\prod_{i=0}^{N-1} \bP_{R} \Psi_{R} (x_{i}) > =  det_{0}( i\slp + \lnA)
\prod_{i,j =0 i\neq j}^{n_{R}-1} (\mid x_{i} - x_{j} \mid )
^{\frac{-\alpha_{i}\alpha_{j}}{(2 \pi)^{2}}} \label{29}
\eeq

\hs This result is in full agreement with the one previously obtained in 
\cite{BC}.\\
\hs The computation of non-minimal functions is, though tedious, straightforward.
The interested reader will find guiding lines in \cite{mnt}.\\

\vspace{0.2cm}

\hs To conclude this letter, as promised in the introductory paragraph, 
we now go back to trivial topology and consider massive fermions. To be more
specific let us examine the following generating functional:

\beq
Z= \int D\bP ~ D\Psi  \exp \{-\int d^2x~ [\bP (i\slp + m + \slA )\Psi ]\}
\label{30}
\eeq

\hs Following the steps of the previous case we make a gauge transformation
in the fermionic variables and represent $\delta(\partial_{\mu}j_{\mu})$
as in (\ref{8}). Calling $B'_{\mu}=B_{\mu}+A_{\mu}$ we arrive at

\beqn
Z & = &  \int D\bP ~ D\Psi  DB'_{\mu} D\theta
\exp \{-\int d^2x~ [\bP i\slp \Psi   
 +  j_{\mu} B'_{\mu} + m \bP \Psi \nonumber\\
& + & \frac{1}{\pi} B'_{\mu} \epsilon_{\mu\nu} \partial_{\nu} \theta 
 -  \frac{1}{\pi} A_{\mu} \epsilon_{\mu\nu} \partial_{\nu} \theta] \} 
\label{31}                                                                          
\eeqn

We make now a chiral change with parameter $\sigma$ in the fermionic 
variables:

\beqn
Z & = & \int D\bP ~ D\Psi  DB'_{\mu} D\theta
\exp \{-\int d^2x~ [\bP i\slp \Psi   
 +  j_{\mu} B'_{\mu} + m e^{2\gamma_{5}\sigma} \bP \Psi \nonumber\\
 & + & \frac{1}{\pi} B'_{\mu} \epsilon_{\mu\nu} \partial_{\nu} \theta 
 -  \frac{1}{\pi} A_{\mu} \epsilon_{\mu\nu} \partial_{\nu} \theta  
-\frac{1}{\pi} B'_{\mu} \epsilon_{\mu\nu} \partial_{\nu} \sigma + 
\frac{1}{2\pi} \sigma \Box \sigma] \}
\label{32}                                                                          
\eeqn

Again in order to eliminate the linear term in $B'_{\mu}$ we identify
$\sigma=\theta$. We then find

\beqn
Z[A_{\mu}] & = & \int D\bC ~ D\chi DB'_{\mu} D \sigma_{\mu}  
\exp \{-\int  d^2x [\bC ~(i\slp + m e^{2 \gamma_{5} \sigma } + \slB ')\chi 
+ \nonumber\\
& + & \frac{1}{\pi}
(\epsilon_{\mu\nu} \partial_{\nu}
\sigma A_{\mu} + 
\frac{1}{2} \sigma \Box \sigma )] \}   \label{33}
\eeqn

\hs At this point we shall make a perturbative expansion in the fermionic 
mass $m$. Obviously, the building block of this expansion is the following 
object:

\beq
\bC (x) ~ e^{2 \gamma_{5} \sigma (x)} \chi (x) = \chi^{ \dagger}_{2}~ 
e^{2 \sigma (x)} \chi_{1} +
\chi^{ \dagger}_{1}~ e^{-2 \sigma (x) }\chi_{2} \label{34}
\eeq

\hs Using this explicit expression, $Z_{m}[A_{\mu}]$  can be readily written 
in terms of fermionic and bosonic v.e.v's in the form:

\beqn
Z_{m}[A_{\mu}] & = & Z_{m=0}^{Fer}[coset] Z[A_{\mu}]^{Bos} \nonumber \\
& \sum_{j=0}^{\infty} & \frac{m^{2j}}{(j!)^{2}} \int \prod_{k=1}^{j} d^2x_{k} 
d^2y_{k} 
<\chi^{ \dagger}_{2}~ \chi_{1}(x_{k}) \chi^{ \dagger}_{1}~ 
\chi_{2}(y_{k})>_{Fer. Coset} \nonumber \\
< & \prod_{k=1}^{j} & e^{2[\sigma (x_{k}) - \sigma(y_{k})]} >_{Bos.\sigma, A_{\mu}} 
\label{35}
\eeqn

\hs Then, exactly as it happens in the massless case (Eq.(\ref{2})) the fermionic coset 
partition function can be extracted as an overall factor, but no complete 
factorization takes place, since every term in the mass expansion contains
a fermionic v.e.v.. However, if one sets 
$A_{\mu}=0 $ the r.h.s. of Eq.(\ref{30}) becomes the partition function of a bosonic 
Sine-Gordon model, as expected (Ref.\cite{Col},\cite{Na}). This fact can be 
easily verified just by 
evaluating the fermionic factor in the series, which can be done in a simple way using, 
for instance, the standard decoupling technique (See Ref.\cite{Ga}).\\

In summary we have extended the factored coset approach to bosonization proposed
in (Ref.\cite{Ni}), in two directions. Firstly we discussed the case of massless
fermions coupled to an Abelian gauge field with non-zero topological charge. 
We were able to show that the coset factorization takes place exactly as in
the $N=0$ case, but with both $Z_{Fer}$ and $Z_{Bos}$ containing the topological
structure, i.e. the corresponding actions are dependent on $N$. Finally we
considered the case $N=0$ but with massive fermions. As it is well-known, the
massive determinant cannot be exactly solved due to the chiral non-invariance
of the mass term. This fact led us to make a perturbative expansion in the mass.
Concerning this case our main conclusion is that no complete factorization is
obtained because the constrained fermionic action enters the game through
the v.e.v.'s which are present in the perturbative series of eq.(\ref{35}).

\vspace{1cm}
\noindent Acknowledgement\\
We thank Fundaci\'on Antorchas for financial support.\\

\end{document}